# Quantitative Morphological and Biochemical Studies on Human Downy Hairs using 3-D Quantitative Phase Imaging

**SangYun Lee[a], Kyoohyun Kim[a], Yuhyun Lee[b], Sungjin Park[b], Heejae Shin[b], Jongwon Yang[b], Kwanhong Ko[b], HyunJoo Park[a], and YongKeun Park[a]**
[a]Korea Advanced Institute of Science and Technology, Department of Physics, Daejeon 305-701, Republic of Korea.
[b]Daejeon Dongsin Science High School, Daejeon 300-310, Republic of Korea.

**Abstract.** This study presents the morphological and biochemical findings on human downy arm hairs using 3-D quantitative phase imaging techniques. 3-D refractive index tomograms and high-resolution 2-D synthetic aperture images of individual downy arm hairs were measured using a Mach-Zehnder laser interferometric microscopy equipped with a two-axis galvanometer mirror. From the measured quantitative images, the biochemical and morphological parameters of downy hairs were non-invasively quantified including the mean refractive index, volume, cylinder, and effective radius of individual hairs. In addition, the effects of hydrogen peroxide on individual downy hairs were investigated.

**Keywords:** body hair, quantitative phase imaging, synthetic aperture imaging, optical diffraction tomography.

**Address all correspondence to:** YongKeun Park, Korea Advanced Institute of Science and Technology, Department of Physics, 291 Daehak-Ro Yusung-Gu, Daejeon 305-701, Republic of Korea. Tel: (82) 42-350-2514; Fax: (82) 42-350-7160; E-mail: yk.park@kaist.ac.kr

## 1. Introduction

Human body hairs exhibit distinct morphologies at their relative sites in the body, from soft downy hairs on the arms to long stiff hairs on the head. The characteristics of different hairs have been regarded as one of the evolutionary consequences incurred by human bipedalism (1, 2). While human body hairs have lost their thermoregulatory roles in maintaining warmth through an evolutionary process, decorative aspects of hairs have gained attentions and led many researchers to investigate the microstructures of downy hairs using diverse microscopic techniques, including atomic force microscopy (3-5), scanning electron microscopy (6-8), transmission electron microscopy (9, 10), confocal microscopy (11, 12) and infrared spectroscopy (13-15) under various experimental conditions. Recently, chemical analyses of hairs have been extensively carried out for drug detection (e.g. alcohol and cocaine), particularly in the fields of forensic science (16-20). However, previous approaches cannot provide non-invasiveness and quantitative imaging capabilities. Furthermore, most previous studies have used expensive instrumentation, which prevents these imaging techniques from being utilized in general laboratories in which body hairs are studied.

Here, we present quantitative phase imaging (QPI) as an effective imaging tool to study the morphological and biochemical properties of downy hairs in a non-invasive and quantitative manner. QPI techniques provide quantitative measurements of optical phase delay introduced by intrinsic refractive index (RI) distributions of transparent cells using interferometry (21, 22). QPI techniques have been applied previously for biological studies of cells and tissues including red blood cells (23-27), cell growth monitoring (28), neurons (29), and optical imaging of tissue slices (30). The feasibility of non-invasive and quantitative imaging of QPI were demonstrated with individual downy hairs using the QPI techniques. The high-resolution 2-D holographic synthetic aperture images and 3-D RI distribution maps of individual hairs were measured, from which the morphological and biochemical properties were retrieved, including the mean RI, volume, cylinder, and effective radius of individual hairs. Furthermore, we investigated the effects of hydrogen peroxide ($H_2O_2$), which is one of the widely-used hair bleaching agents, on individual downy hairs using the present method.

## 2. Materials and Methods

### 2.1 Sample Preparations

A total 13 human downy hairs were collected from one arm of a healthy donor (Asian male) [Fig. 1(a1)]. Each hair was gently posed on the top of a coverslip (24 × 50 mm, C024501, Matsunami, LTD, Japan) with oil immersion ($n = 1.518$), which decreases the RI contrast between the hairs and the medium. The sample was then covered with another coverslip [Fig. 1(a2)]. To study the effects of hydrogen peroxide on the hair structures, four hair samples were treated with 3% $H_2O_2$ solution (Sigma-Aldrich, St. Louis, Missouri, US) for 24 hours. The shape of the edges of the hairs cells was imaged before and after the $H_2O_2$ treatments.

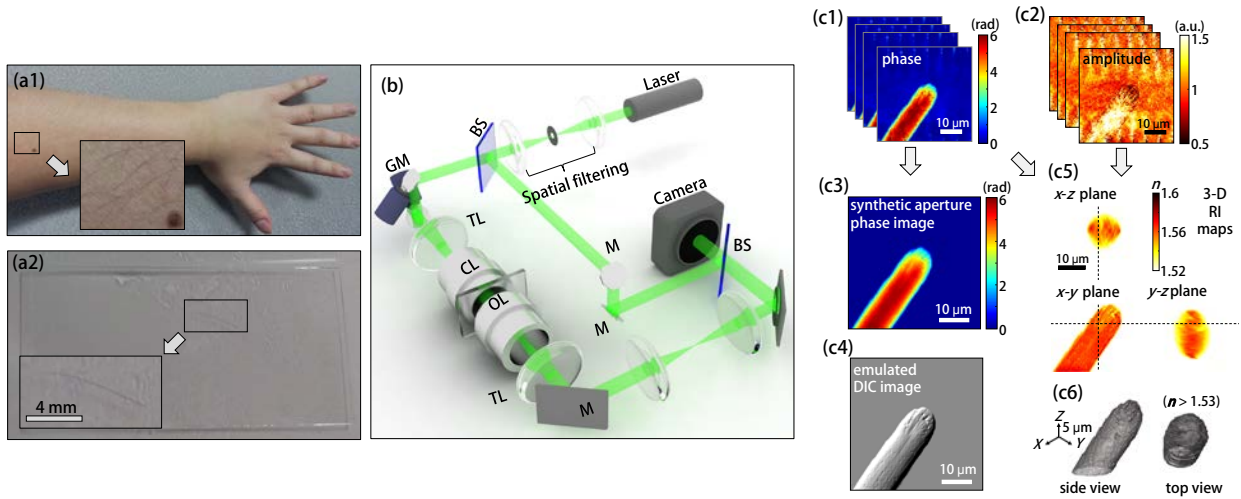

**Fig. 1** (a) Human downy hair preparation. (a1) Left arm of a healthy donor from which downy hairs were collected. (a2) A downy hair with immersion oil loaded between two coverslips. (b) Mach-Zehnder interferometric microscopy equipped with a two-axis galvanometer mirror. BS: beam splitter; GM: galvanometer mirror; CL: condenser lens; OL: objective lens, M: mirror. (c) Holographic image reconstruction process. A set of retrieved (c1) phase and (c2) amplitude maps of complex optical fields. (c3) Synthetic aperture phase image. (c4) Emulated DIC image. (c5) Reconstructed 3-D RI distributions of a hair sample. (c6) 3-D RI isosurface of a hair tomogram.

### 2.2 Optical Setup for QPI

For quantitative measurements of human downy hairs, we employed a Mach-Zehnder interferometric microscope equipped with a two-axis galvanometer mirror (31-33). The schematic of the setup is shown in Fig. 1(b). A diode-pumped solid state laser ($\lambda$ = 532 nm, 50 mW, Cobolt Co., Solna, Sweden) was used as a coherent light source. A beam splitter (BS, BS016, Thorlabs, U.S.A.) divides a laser beam into two arms: a reference beam and a sample beam. Sample beam passes through a downy hair loaded on the sample stage of an inverted microscope. A two-axis galvanometer mirror (GM, GVS012/M, Thorlabs, U.S.A.) varies the angle of the illumination beam impinging onto the hair sample, from which 2-D optical field images of the sample are obtained with various illumination angles. For the reconstruction of off-axis interference patterns, the reference and the sample beams are recombined with a tilted angle by another beam splitter, and the resultant interferograms of samples are recorded by a high-speed CMOS camera (Neo sCMOS, Andor Inc.,

Northern Ireland, U.K.).

An objective lens [UPLFLN, 60×, Numerical aperture (NA) = 0.9, Olympus Inc., San Diego, C.A., U.S.A.] was used as a condenser lens with the tube lens of a focal length of 200 mm. For the imaging purpose, a high-N.A. objective lens (PLAPON, 60×, Oil immersion, NA = 1.42, Olympus Inc., San Diego, C.A., U.S.A.) was used with an additional telescopic 4-*f* system, and the total magnification of the imaging system is 250×. The camera has 1,776 × 1,760 pixels with a pixel size of 6.5 μm. The field of view at the sample plane was 46.18 × 45.76 μm$^2$. Each hair sample was illuminated with plane waves at 300 different incidence angles, which were systematically controlled by the two-axis GM at a frame rate of 100 Hz.

*2.3 Image Reconstruction Procedures*

Complex optical fields of the sample were retrieved from the interferogram recorded by the QPI technique, via a retrieval algorithm based on Fourier transform (34, 35). The phase and amplitude maps of a representative hair are shown in Figs. 1(c1) & 1(c2).

From a set of retrieved phase images [Fig. 1(c1)] with various illumination angles, the 2-D high-resolution synthetic aperture phase image [Fig. 1(c3)] was constructed with a synthetic aperture imaging algorithm (36, 37). By numerically extending the aperture size, the synthetic aperture algorithm fully used the high spatial frequency information of a sample which cannot be accessed with just a single laser illumination angle. The numerical extension of an aperture was conducted in the 2-D Fourier space, and the resultant 2-D synthetic aperture phase image [Fig. 1(c3)] exhibited a higher spatial resolution and signal-to-noise ratio (SNR), compared to the phase image from a single hair interferogram [Fig. 1(c1)] (38, 39).

In principle, the resolution of an imaging system is determined by the NAs of an objective lens. In general, for 2-D QPI with a single illumination, the maximum accessible spatial frequency $|k_{max}|$ is determined as $2\pi NA_{imag}/\lambda$, where $NA_{imag}$ represents the NA of the imaging system. In synthetic aperture imaging, the maximum spatial frequency can be further extended by combining multiple phase images obtained with various incident angles; the maximum spatial frequency is extended to $2\pi(NA_{imag}+NA_{illum})/\lambda$, where $NA_{illum}$ is the N.A. of the condenser lens. The SNR of the synthetic aperture phase image is also more significantly enhanced than that of a single phase image, mainly due to speckle noise reduction.

A differential interference contrast (DIC) image can also be numerically reconstructed from the synthetic aperture phase image [Fig. 1(c4)]. This emulated DIC image is readily obtained from a synthetic aperture phase image by numerically interfering an original phase image and a slightly translated synthetic aperture phase image with an additional phase shift (40). Because the optical imaging contrast in DIC images is a result of the phase gradient, the imaging contrast is enhanced for objects with high RI changes or gradients, such as subcellular organelles in unstained biological cells or defects in non-biological samples.

To reconstruct the 3-D RI distribution of a hair sample from a set of retrieved complex optical fields, we used the optical diffraction tomography (ODT) algorithm (32, 33, 41). Compared to the projection algorithm (42, 43), the ODT algorithm considers light diffraction at samples, and thus provides high-resolution tomographic reconstruction with better image qualities especially for large samples with high RI contrast (32).

*2.4 Statistical Analysis*

*P*-values are calculated by two-tailed paired t-test comparing quantitative parameters of human arm hairs. All the numbers follow the ± sign in the text are standard deviations.

**3. Result and Discussion**

*3.1 High-resolution 2-D Phase Images and 3-D RI Maps of Human Downy Hairs*

In order to measure the morphologies of individual downy hairs, we measured and took 3-D RI maps and 2-D synthetic aperture phase images of human downy hairs collected from arms, using the Mach-Zehnder laser interferometry equipped with a two-axis galvanometer mirror (See *Materials and Methods*). Figure 2 shows the measured images of representative human downy hairs. Measured hair samples were categorized into two groups based on their morphologies; hairs in the soft type exhibit smooth surfaces and internal RI distributions [Figs. 2(a1−a3)], and hairs in the rough type present complex internal structural defects and spiky surfaces [Figs. 2(b1−b3)].

High-resolution 2-D synthetic aperture phase images [Figs. 2(a1) and (b1), top], corresponding emulated DIC images [Figs. 2(a1) and (b1), bottom] and three cross-sectional slices of the reconstructed 3-D RI tomograms [Figs. 2(a2) and (b2)] of the representative downy hairs showed significant morphological differences (See Materials and Methods). The renderings of 3-D RI isosurfaces ($n > 1.545$) of the hair samples are also shown in Figs. 2(a3) and (b3).

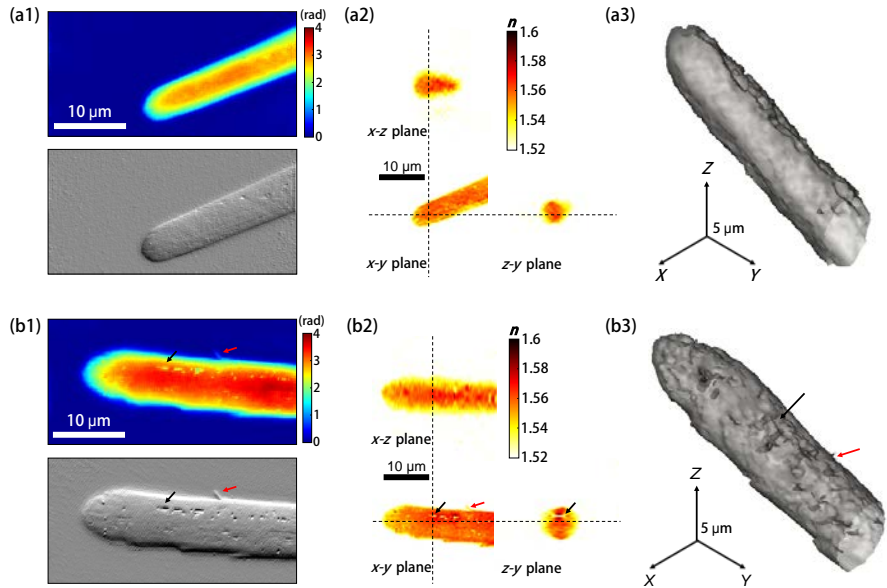

**Fig. 2** Quantitative phase images of representative human downy hairs for soft and rough type ((a) and (b), respectively): (a1) and (b1), the 2-D synthetic aperture phase image (top) and the corresponding emulated DIC image (bottom) of the representative hair samples. (a2) and (b2), the cross-sectional slices of the reconstructed 3-D RI tomogram in *x−y*, *z−y*, and *x−z* plane. (a3) and (b3), perspective view of the representative 3-D hair RI isosurface. RI threshold set as a value of 1.545. Each colored arrow in the figures represents the same fine structures of the measured downy hair.

In total, 13 human downy hairs were measured; 6 hair cells were categorized as the soft type, and 7 hair cells were in the rough type. The representative holographic images of the soft type hair [Figs. 2(a1−a3)] had soft and continuously connected cell boundaries without any apparent structural defects. The relatively homogenous RI distribution of the soft type hair [Fig. 2(a2)] reflects the homogeneity of its internal structures. Figures 2(b1−b3) show quantitative phase images of the representative rough type downy hair.

Significantly low speckle noise level in the backgrounds of the DIC image indicates that the observed fine structures in the hair sample are from real structural defects. The characteristic fine structures, indicated with the colored arrows in Fig. (b1), are also shown in the reconstructed 3-D RI tomogram [Fig. 2(b2)]. Particularly, the structural defect [the black arrow in Fig. 2(b1)] can be clearly seen in the *z − y* sectional RI slice. Finally, the RI isosurface of the rough type hair [Fig. 2(b3)] shows more complicated surface structures when compared to the RI isosurface of the soft type hair [Fig. 2(a3)].

*3.2 Retrieval of Quantitative Downy Hair Parameters*

To demonstrate the quantitative imaging capability of the present method for the study of downy hairs, we retrieved the morphological and biochemical parameters from measured 3-D RI maps of individual hairs. The retrieved parameters included mean RI values, the volumes for 20-μm-length hairs [Fig. 3(a)], cylindrical radii of the hairs as a function of length [Fig. 3(b)], and the effective radii of hair edges [Fig 3(c)].

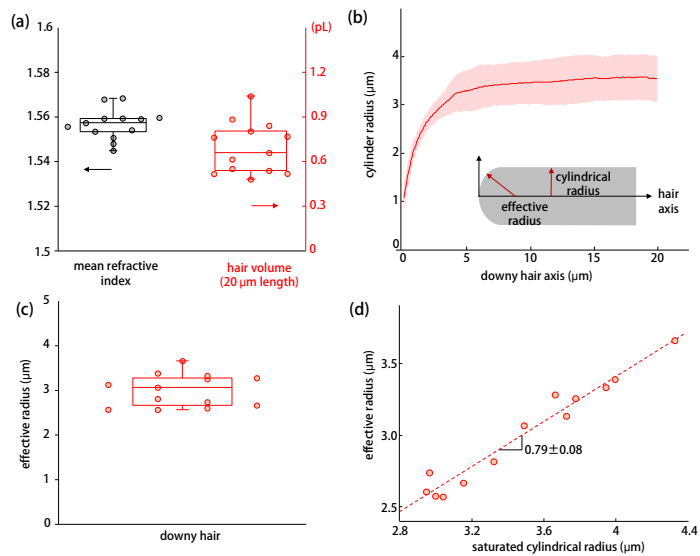

**Fig. 3** Biochemical and morphological parameters of human downy hairs. (a) Retrieved mean RIs and volumes of 20-μm-length hairs. (b) Cylindrical radii of hairs as a function of length. The shaded area represents standard deviations. Inset: a schematic describing the radius and effective radius of a hair. (c) Retrieved effective radius. (d) Correlation map between the effective radii and the cylindrical radii. Each colored dot denotes individual hairs. Boxes, median with upper and lower quartiles; Whiskers, parameter range.

To obtain the mean RI value of a hair, we averaged the RI values over the sample area in the focal plane [i.e. the $x - y$ plane in Figs. 2(a2) and (b2)]. Volumes of the 20-μm-length hairs were calculated by integrating voxels higher than the RI threshold ($n = 1.53$). The mean values of the retrieved mean RI and volume for the 20 μm length hairs were $1.557 \pm 0.007$ (mean ± std) and $692 \pm 173$ fL, respectively, which are in good agreement with a previous RI measurement (44).

The cylindrical radii of hairs as a function of hair lengths are retrieved assuming a cylindrical symmetry [*inset* in Fig. 3(b)]. Then, the cylinder radius represents the hair thickness as a function of hair length. The graph in Fig. 3(b) depicts the measured cylindrical radii for 13 downy hairs along their hair axis. The red line denotes the mean cylinder radii, and the shaded region corresponds to standard deviations. The mean cylindrical radii seems to saturate beyond a length of 5 μm.

The roundness of the hair tips can be effectively described by measuring the effective radii of the tips [Fig. 3(c)]. In order to retrieve the effective radius of a hair $R$, the following relation was used: $R = (r^2 + h^2)/2h$, where $r$ is the radius of a circular slice orthogonal to the hair axis, $h$ is the distance between the circular slice and the hair tip. For the case of a perfect sphere, any circular slice of the sphere with a different $h$ gives the same $R$. This is not the case for hairs, and the height $h$ should be determined appropriately in order to have a sphere with an effective radius $R_{eff}$ describing the roundness of the hair edge. We determined $h$ from the plane on the hair axis where the cylindrical radius equals 85% of the saturated cylindrical radius. Results are shown in Fig. 3(c). The mean effective radius of the measured hairs was $3.00 \pm 0.37$ µm.

To address the relationship between the effective radii $R_{eff}$ and the saturated cylindrical radii of the hairs $r_{sat}$, the correlation of these two parameters were investigated [Fig. 3(d)]. The correlation clearly shows a linear relationship; they are well described by a linear regression model $R_{eff} = 0.79\ r_{sat} + 0.26$ µm with $R^2 = 0.96$. This linear relation of these radii implies that each human arm hair shares a morphological similarity with each other. In addition, the coefficient of proportionality less than one means that hair edges are sharper than ideal hemispheres, in accordance with the measured 3-D RI tomograms.

### 3.3 Effects of Hydrogen Peroxide on Individual Human Downy Hairs

To further show the capability of QPI, we performed experiments to study the effects of hydrogen peroxide on individual hairs. Using the present method, four hairs were systematically measured before and after a 24-hour treatment with a 3% hydrogen peroxide solution (See *Methods*). Hydrogen peroxide is one of the widely used bleaching agents which induces irreversible alterations in the physicochemical properties of the melanin granules in hair proteins (45). It is also known that physical damage occurs to keratin proteins from oxidization.

To investigate possible alterations induced by hydrogen peroxide, we measured the morphological and biochemical properties of hairs. The $x$–$y$ cross-sectional slices of the reconstructed 3-D RI tomograms for four intact arm hairs [Figs. 4(a1–a4)] and the same hairs treated with 3% $H_2O_2$ [Figs. 4(a5–a8)] are shown in Fig. 4(a). The arrows and the dashed region denote the same fine structures, which were used to track the same hairs before and after the $H_2O_2$ treatment. Each graph in Fig. 4(b) and Fig. 4(c), respectively, describes the mean RIs and volumes of the 20-µm-length hairs before and after $H_2O_2$ treatment. To retrieve these parameters, the same procedures described above were used. The mean values of RI for intact and $H_2O_2$ treated hairs were $1.557 \pm 0.003$ and $1.556 \pm 0.005$, respectively. In addition, the mean volume for the intact and $H_2O_2$ treated hair groups was $724 \pm 141$ and $731 \pm 166$ fL, respectively. The paired $t$-test yielded $p$-values of 0.7330 for the mean RI and 0.7217 for the volume between the two hair groups, and thus does not show statistical difference before and after the treatment.

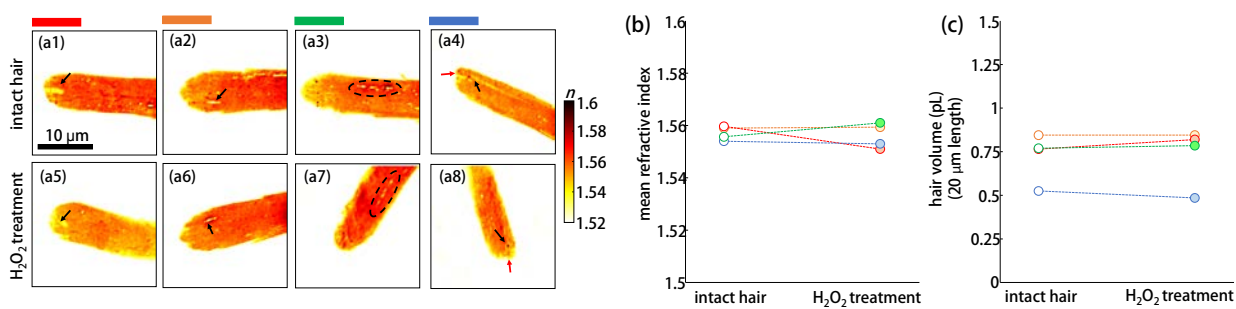

**Fig. 4** Effects of hydrogen peroxide on human downy hairs. (a) $x$–$y$ cross-sectional slices of RI tomograms for (a1–a4) intact hairs and (a5–a8) the same hairs after a 24 hr $H_2O_2$ treatment, respectively. (b) Mean RIs and (c) volumes of four hairs before and after the treatment. Same color denotes the same individual downy hair.

The statistical indifference between the intact hairs and the $H_2O_2$ treated ones, however, does not imply that the edge structures of human downy hairs are not affected by hydrogen peroxide. It seems that alterations in the edge structures of body hairs by hydrogen peroxide are small as predicted by minute morphological changes in Figs. 4(a1–a8), and so comparable with the measurement errors. Noticeably, we could not observe black stains maybe responsible for melanin pigments at hair edges, contrary to observations at thicker parts of the hair. The absence of melanin pigments at the hair edges possibly give rise to minute changes in the measured parameters by hydrogen peroxide.

## 4. Conclusion

Herein, we performed quantitative and non-invasive optical measurements on human downy arm hairs using 3-D quantitative phase imaging. With a Mach-Zehnder interferometer equipped with a dual axis galvanometer mirror, 3-D RI tomograms and 2-D synthetic aperture images of individual hairs were measured. To fully exploit the quantitative imaging capability of the present method, biochemical and morphological parameters including mean RI, volume, cylinder radius, and effective radius for individual downy hairs were retrieved from the measured 3-D RI maps. Finally, the compositional and structural alterations in the downy hairs by hydrogen peroxide were also investigated with the present method.

We expect that the unique advantages of the QPI techniques with its high-resolution 3-D RI imaging capabilities can be beneficial in investigating alterations in human hairs from various chemical agents such as bleaching agents (9, 13), shampoos, and rinses. Furthermore, structural variations in hair between different human species (6, 46) or gathered body sites (47, 48) can also be investigated from both compositional and morphological aspects. Furthermore, recently advanced QPI techniques including a readily implementable QPI unit (49, 50), real-time 3-D measurements (51), spectroscopic RI imaging (52-55), polarization-sensitive imaging (56, 57), and a super-resolution technique (58) will also extend the applicability toward quantitative studies on the dynamic phenomena relevant to hair growths (59, 60) or formations of structural defects under diverse experimental conditions.

## Acknowledgements

This work was supported by KAIST-Khalifa University Project, APCTP, the Korean Ministry of Education, Science and Technology, and the National Research Foundation (2012R1A1A1009082, 2014K1A3A1A09063027, 2013M3C1A3063046, 2012-M3C1A1-048860, 2014M3C1A3052537).